# A fully differentiable ligand pose optimization framework guided by deep learning and traditional scoring functions

Zechen Wang[1#], Liangzhen Zheng[23#*], Sheng Wang[3], Mingzhi Lin[3], Zhihao Wang[1], Adams Wai-Kin Kong[4], Yuguang Mu[5], Yanjie Wei[2*], Weifeng Li[1*]

1. School of Physics, Shandong University, Jinan, Shandong 250100, China
2. Shenzhen Institute of Advanced Technology, Chinese Academy of Sciences, Shenzhen, Guangdong 518055, China
3. Shanghai Zelixir Biotech Company Ltd., Shanghai 200030, China
4. Rolls-Royce Coporate Lab, Nanyang Technological University, 637551, Singapore
5. School of Biological Sciences, Nanyang Technological University, 637551, Singapore

[#]The authors contribute to the work equally.
*Corresponding Authors.
E-mail: astrozheng@gmail.com (L. Zheng), yj.wei@siat.ac.cn (Y. Wei) and lwf@sdu.edu.cn (W. Li)

## ABSTRACT

The machine learning (ML) and deep learning (DL) techniques are widely recognized to be powerful tools for virtual drug screening. The recently reported ML- or DL-based scoring functions have shown exciting performance in predicting protein-ligand binding affinities with fruitful application prospects. However, the differentiation between highly similar ligand conformations, including the native binding pose (the global energy minimum state), remains challenging which could greatly enhance the docking. In this work, we propose a fully differentiable framework for ligand pose optimization based on a hybrid scoring function (SF) combined with a multi-layer perceptron (DeepRMSD) and the traditional AutoDock Vina SF. The DeepRMSD+Vina, which combines (1) the root mean square deviation (RMSD) of the docking pose with respect to the native pose and (2) the AutoDock Vina score, is fully differentiable thus is capable of optimizing the ligand binding pose to the energy-lowest conformation. Evaluated by the CASF-2016 docking power dataset, the DeepRMSD+Vina reaches a success rate of 95.4%, which is by far the best reported SF to date. Based on this SF, an end-to-end ligand pose optimization framework was implemented to improve the docking pose quality. We demonstrated that this method significantly improves the docking success rate (by 15%) in redocking and crossdocking tasks, revealing the high potentialities of this framework in drug design and discovery.

# 1. Introduction

Computer technology is playing an increasingly important role in drug design[1, 2]. An important task in computer-aided drug design is to discover lead compounds with high binding affinity to disease-causing proteins of interest[3, 4, 5, 6]. Since binding affinity prediction is heavily dependent on the position of the ligand in the binding pocket, it is critical for the most favorable pose (with the lowest binding free energy) to be easily singled out from among numerous pose decoys[7]. Binding pose discrimination in turn directly affects the confidence of virtual screening results.

In structure-based virtual screening, molecular docking programs are often used to explore potential conformations of ligands upon binding[8, 9, 10, 11, 12]. Molecular docking consists of two basic steps: conformation sampling and scoring[13, 14]. Conformation sampling refers to searching for potential binding poses of the ligand at the pocket of the target[15]. The Monte Carlo algorithm is one popular sampling method that is employed to sample conformations that satisfy physicochemical laws within a defined space. Scoring refers to predicting the binding affinity between the protein and the sampled conformation of the ligand. Then, the sampled conformations are sorted by predicted binding affinity, and one or several of them with greatest binding affinity are selected as the potential docking poses to represent protein-ligand interaction patterns. Conformation sampling in current docking programs generally works well, but the accuracy of the SFs is unsatisfactory, which is still the main factor that limits the performance of molecular docking[2, 16, 17]. Most of the SFs in docking programs are based on classical SFs, which are generally classified into three categories: physics-based, empirical, and knowledge-based. Such SFs were usually pre-designed by experts, and parameters were determined by fitting a small number of protein-ligand complex structures[16, 18]. However, each SF has obvious shortcomings which limit the performance of the docking program[12]. For example, although physics-based SFs work well for affinity prediction, this method is usually time-consuming[19]; empirical SFs are more efficient than physics-based SFs, but their prediction accuracy is limited[20]; knowledge-based SFs can balance accuracy and efficiency, but it is difficult for them to determine the reference state[12, 21].

In the past decade, ML and DL algorithms have been successfully applied to construct new SFs. Compared with classical SFs, ML- and DL-based SFs are allowed to learn the protein-ligand interaction information from larger training datasets[17, 22, 23]. In addition, both ML and DL are better at dealing with the non-linear relationship between features and labels, contributing to the increasing popularity of their implementation for SFs[24]. Most of the published popular ML- and DL-based SFs, such as RF-Score[25], NNScore[26], $K_{deep}$[27], AGL[28], OnionNet[29], OnionNet-2[30], OnionNet-SFCT[31], HPC/HWPC-GBT[32] and OPRC-GBT[33], etc., have been shown to have strong power in protein-ligand binding affinity prediction. However, the ability of such SFs to select near-native conformations from a pool of docking poses generated by docking programs (referred to as docking power) has rarely been tested. $G_{NINA}$ is a computational docking software that uses 3D CNN models to rescore docking poses[15]. Furthermore, studies have shown that there is no strong correlation between the scoring power and docking power of SFs[34]. For instance, while some ML-based SFs demonstrate superior scoring power, they perform poorly in docking power, and may in fact perform worse than classical SFs[35]. Therefore, an ideal SF would be capable of selecting near-native conformations while simultaneously being able to explicitly guide the sampling process in molecular docking scenarios.

The performance of DL-based SFs depends largely on the characterization of protein-ligand interactions[17, 36]. It is well known that majority of protein-ligand recognitions are determined by the synergy of multiple intermolecular non-bonded interactions, such as hydrogen bonding, hydrophobic interactions, electrostatics, π-π stacking and van der Waals interactions, etc.[37, 38, 39] What these interactions have in common is that they mainly depend on the distance and angles between the interacting atoms or functionals groups.

In this work, we characterized protein-ligand interactions based on physical features rooted in van der Waals and electrostatic interactions, and constructed an efficient MLP model, DeepRMSD, for predicting the RMSD of docking poses relative to the native pose of the ligand. Taking the CASF-2016 docking poses as the test set, DeepRMSD shows high accuracy rate in the RMSD range above 0-4 Å, while AutoDock Vina performs well in the RMSD range below 0-3 Å. We combine DeepRMSD with Vina score to get a new SF, called DeepRMSD+Vina. In the docking power test of CASF-2016 benchmark, DeepRMSD+Vina achieved the highest success rate of Top 1 of 95.4 % (the percentage of targets where the RMSD of the top-ranked pose is less than 2 Å), followed by AutoDock Vina with 90.2 % of the rate of Top 1.

DL-based SFs for molecular docking are not only used to rescore poses generated by other docking programs, but also should be adopted to optimize the ligand conformations. Recently, a conformation optimization method based on a geometric deep learning approach has been proposed, but it has not been tested in practical application scenarios[40].

In view of the superiority of the DeepRMSD+Vina score in docking power, we further constructed a fully differentiable docking pose optimization framework. In this framework, the sub-molecular translational and rotational movements are guided by back-propagation gradients of the SF.

The optimization algorithm on docking poses is evaluated using CASF-2016 as a benchmark. The results show that the optimization success rate of our method in the low RMSD interval (0-3 Å) is 69 %. In addition, our framework is tested on the practical redocking and cross-docking tasks in which improved docking success rates are demonstrated. Our DL framework that integrates the prediction and optimization of poses of ligands will greatly improve the quality of the poses output in molecular docking, and thus has significant practical application value. The DeepRMSD+Vina model and the optimization framework can be available at public GitHub repository https://github.com/zchwang/DeepRMSD-Vina_Optimization.

## 2. Methods
### 2.1 Datasets

The training, validation, and test sets included the native protein-ligand complex structures in PDBbind database v.2019 (composed of the general set and the refine set, totaling 17652 native protein-ligand complexes) and the docking poses generated. The division of the dataset was consistent with previous work[30] and it was ensured that there were no overlapping samples between the training, validation, and test sets. In CASF-2016 benchmark[34], about 100 docking poses were generated for each protein-ligand complex to evaluate the docking power of SFs, with the RMSD value ranging from 0 to 10 Å; these docking poses were employed to evaluate the SF we propose in this work.

We excluded samples with short peptides as ligands because they usually have more rotatable bonds. In addition, the samples that could not be processed normally by our algorithm were not taken in account. We used AutoDock Vina to re-dock the protein-ligand pair included in the general set and refined sets but not the CASF-2016 core set to generate enough docking poses for training. Each protein-ligand pair was docked 10 times, and 10 docking poses were generated each time. The sampling space was a $15\text{Å} \times 15\text{Å} \times 15$ Å cubic, which was centered on the geometric center of the native binding pose of the ligand. The RMSD value between the docking pose and the native conformation was calculated by Hungarian algorithm embedded in spyrmsd[41]. The RMSD distribution of docking poses in the training set is shown in Supplementary Fig. 3.

We also evaluated our algorithm in two practical docking scenarios: redocking and cross-docking tasks. We constructed the redocking set by re-docking a ligand to its receptor in the CASF-2016 core set with AutoDock Vina; each native protein-ligand complex had no more than 20 docking poses. For the cross-docking task, a cross-docking dataset called 3D DISCO was employed, which contains 4399 protein-ligand complexes across 95 protein targets, with an average of 46 ligands for each target[42]. These targets were derived from the DUD-E database to ensure that a wide range of protein families were included in the cross-docking dataset. Unlike redocking, in the cross-docking processes, a ligand was extracted from a co-crystal structure and then was docked to another crystal structure of the same target. Obviously, this task is more challenging, but is more faithful to a real virtual screening scenario where the aim is to find new hits to bind target structures that are typically co-crystallized with some known ligands. The RMSD number distributions of docking poses in the redocking and cross-docking sets are shown in Supplementary Fig. 4a and 4b.

**2.2 Descriptors**

It is well known that protein-ligand binding is mainly affected by non-bonded interactions, which comprise two types, namely van der Waals interactions and electrostatic interactions. In light of this, we extracted protein-ligand interaction features from these two basic interactions, and then used a MLP to fit the relationship between features and RMSD. The expressions of van der Waals interaction (characterized by Lennard-Jones potential) and electrostatic interaction are shown in Eq. 1 and Eq. 2, respectively.

$$V_{LJ}(r) = \frac{C_{12}}{r^{12}} - \frac{C_6}{r^6} \qquad \text{Eq. 1}$$

$$V_{coul}(r) = \frac{1}{4\pi\varepsilon_0}\frac{q_i q_j}{r} \qquad \text{Eq. 2}$$

Where $C_{12}$ and $C_6$ are the repulsive and attractive parameters respectively, which are usually determined by the types of the two interacting atoms. The $q_i$ and $q_j$ represent the partial charges of two atoms, and $r$ is the distance between two atoms. We aimed to extract the distance and atom type information between the interacting atoms in the protein-ligand complex, and then learn the parameters other than $r$ through the MLP model. As can be seen from Eq.1 and Eq.2, the van der Waals and Coulomb potentials are determined by two parts of parameters, namely the parameters related to $r$ and the non-$r$-dependent parameters determined by the type of atom pairs. Obviously, the $r$-related parameters actually exist in the form of the negative $i$ ($i$ = 1, 6, 12) power of $r$. The $-\frac{1}{r^6}$ term comes from the well-known London dispersion force. However, the $\frac{1}{r^{12}}$ term is a very roughly approximate energy function for the repulsive forces rooted in the Pauli exclusion principle. The repulsive term in the van der Waals interaction can only be effective when the $r$ is small enough (for example, the $r$ is less than the sum of the van der Waals radii of two atoms), and this is indeed a small probability event in protein-ligand binding. In addition, the feature of the negative 12 power of $r$ makes the neural network less robust; that is, a small difference in the $r$ term may cause a large change in the feature value, which may be unfriendly to the neural network. Therefore, the repulsion term ($i$ = 12) was not considered when modeling the protein-ligand interaction in this work. It needs to be emphasized that in the docking pose optimization stage, we added a steric penalty term based on the intermolecular and intramolecular distances to prevent steric collisions between the ligand and the protein and within the ligand molecule.

Input files of the protein and the ligand are in PDBQT format, and files were prepared by the prepare_receptor4.py and prepare_ligand4.py scripts in MGLTools, respectively. The structure information of the protein-ligand complex was obtained from these two input files. First, we classified atoms in the protein into five basic types ($T_r$), namely C, N, O, S, and DU, where DU represents the element types excluded in the first four types. It should be noted that such a simple classification cannot capture differences between atoms with the same element type. In fact, the same element in different residues usually has different physicochemical properties. Furthermore, our previous work also showed that residue types have a positive effect on modeling protein-ligand interactions[30]. Therefore, we further classified the five basic atom types in the protein according to the residue types, such that each residue has five independent residue-atom ($RA$) types. In addition to the 20 residues, we also added an additional type named "OTH" to represent the other groups. Therefore, a total of 105 $RA$ types ($T_{RA}$) were generated.

$$T_{res} \in \{20\ residues, \text{OTH}\}$$
$$T_r \in \{C, N, O, S, DU\}$$
$$T_{RA} \in \{T_{res} \times T_r\}$$

For the atoms in the ligand (L), we defined seven atom types ($T_L$), namely C, N, O,

P, S, HAL, and DU. Here HAL represents the halogen elements (F, Cl, Br and I), and DU represents the element types excluded in first six types, which is consistent with our previous work.

$$T_L \in \{C, N, O, P, S, HAL, DU\}$$
$$HAL \in \{F, Cl, Br, I\}$$

To represent protein-ligand interactions, $105 \times 7$ types of *RA*-atom pairs were applied. The feature calculation formula of each *RA*-atom pair is shown in Eq. 3. The distance $r$ of each *RA*-atom pair was taken from the protein-ligand complex/docking poses. Only *RA*-atom pairs with $r$ within the predefined interaction range contributed to protein-ligand binding. Then, we performed negative $i$ power processing on the distance $r$ of the *RA*-atom pairs that met the distance constraints. Finally, the processed distance parameters $r_{ra,l}^{-i}$ were accumulated according to the specific *RA*-atom type to obtain the final feature of each *RA*-atom type.

$$E(RA, L, i) = \sum_{ra=R, l=L} \frac{1}{r_{ra,l}^i} \quad \text{Eq.3}$$

Where $E(RA, L, i)$ is the feature value of the *RA-L* combination pair. In this work, we set $i = 1$ and $i = 6$ to approximate the Coulomb and the van der Waals interaction terms, respectively. These two types of features were directly concatenated horizontally and then fed into the neural network. Therefore, the interaction information for each protein-ligand complex was encoded in a vector of length 1470.

## 2.3 Model

In this work, we adopted a multi-layer perceptron (MLP) model to construct a scoring function called DeepRMSD for predicting the RMSD of the docking pose with respect to the native pose. The model was constructed and trained with the Pytorch program[43].

Before training, the feature vector of each protein-ligand complex was preprocessed by the Standardization method in the scikit-learn package[44]. Then, the preprocessed 1D vector was fed into the MLP model.

Here, we implemented a simple, lightweight MLP framework, which contained five fully connected layers, with 1024, 512, 256, 128 and 64 neurons respectively, followed by the output layer. In addition, each fully connected layer was augmented by a rectified linear unit (RELU) layer to further improve the nonlinear capability of the model. Here, the stochastic gradient descent (SGD) optimizer was adopted. After careful tests, the learning rate was set to 0.001, the number of samples processed per batch during training was 32 and no dropout layer was added. Details of the hyper-parameter optimization process can be found in Supplementary Table 1. The mean square error (MSE) was used as the loss function. The early stopping strategy was also employed to avoid unnecessary training iterations, and the model that performed best on the validation set was retained.

## 2.4 Docking power

Docking power refers to the ability of a SF to identify near-native conformation from docking poses[34]. Here, the docking poses of each protein-ligand complex were re-scored by DeepRMSD and then were sorted in ascending order by the predicted RMSD value. Ideally, the native binding pose should be identified as the top-ranked one[34]. The docking power was evaluated here based on the percentage of targets where the RMSD of the top-ranked pose was less than 2 Å (Top 1 success rate). Considering that other top-ranked poses generated by molecular docking may be useful as well, the Top 2 and Top 3 success rate were also used in the docking power test. Among the SFs evaluated by CASF-2016, AutoDock Vina performed best, with a Top 1 success rate of 90.2 % in the docking power test. When the CASF-2016 docking poses were re-scored by DeepRMSD alone, the Top 1 success rate was worse than AutoDock Vina, which indicates that DeepRMSD is less sensitive to poses with low RMSD than AutoDock Vina. Interestingly, if the value predicted by DeepRMSD was combined with Vina score (DeepRMSD+Vina), the Top 1 success rate was significantly improved. It should be noted that the performance of DeepRMSD+Vina depends on the weights of its two components, which is optimized according to the Top 1 success rate on CASF-2016 docking poses. It was found that DeepRMSD+Vina performed best in the docking power test when both were equally weighted. For details of the test process, refer to the Supplemental Material.

**2.5 The implementation of inter-molecular energy for AutoDock Vina**

For the Vina score component of the overall score, only the intermolecular interaction term was considered, which was implemented through Pytorch[43]. The intermolecular term of the Vina score is the weighted sum of five terms over all the pairs of atom *i* and atom *j*, which are derived from the protein and the ligand respectively[10, 45]. $d_{ij}$ is defined as the surface distance of the two atoms, which can be calculated by Eq. 9, where $r_{ij}$ is the interatomic distance and $R_{t_i}$ and $R_{t_j}$ are their van der Waals radii. The atom types adopted by AutoDock Vina are consistent with that of the X-score. For this SF, the distance cutoff for all interaction terms is 8 Å. The five terms are shown below (Eq. 4 to Eq. 8), and the weights of the five terms are shown in Supplementary Table 2.

$$gauss_1(d_{ij}) = e^{-(d_{ij}/0.5 Å)^2} \qquad Eq.4$$

$$gauss_2(d_{ij}) = e^{-((d-3Å)/2Å)^2} \qquad Eq.5$$

$$repulsion(d_{ij}) = \begin{cases} d_{ij}^2, & if\, d_{ij} < 0 \\ 0, & if\, d_{ij} \geq 0 \end{cases} \qquad Eq.6$$

$$hydrophobic(d_{ij}) = \begin{cases} 1, & if\, d_{ij} \leq 0.5 \\ 1.5 - d_{ij}, & if\, 0.5 < d_{ij} < 1.5 \\ 0, & if\, d_{ij} \geq 1.5 \end{cases} \qquad Eq.7$$

$$hbonding(d_{ij}) = \begin{cases} 1, & if\ d_{ij} \leq -0.7 \\ \frac{d_{ij}}{-0.7}, & if\ -0.7 < d_{ij} < 0 \\ 0, & if\ d_{ij} \geq 0 \end{cases} \qquad \text{Eq.8}$$

$$d_{ij} = r_{ij} - (R_{t_i} + R_{t_j}) \qquad \text{Eq.9}$$

## 2.6 Optimization

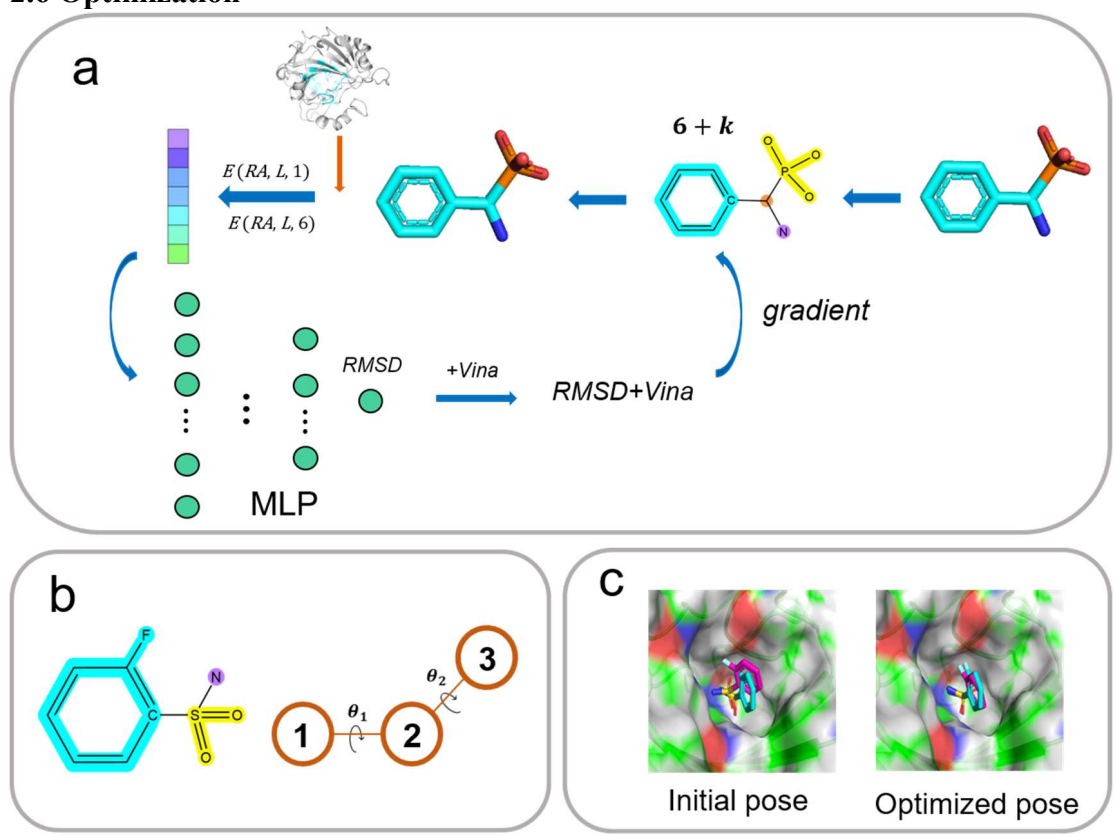

**Figure 1.** Workflow of the DeepRMSD+Vina-based binding pose optimization framework. **a** Overall architecture of the pose optimization framework. **b** Representation of rotatable bonds and subgroups in the ligand. **c** Conformation comparison of a docking pose (pink) with the native pose (cyan) before and after optimization (PDB: 2WEG).

We designed a pose optimization framework based on DeepRMSD+Vina to further improve the quality of poses generated by molecular docking (Fig. 1a). During the optimization, the conformation of the ligand was updated by adjusting the translation, rotation, and torsion of the rotatable bond of the ligand. Here, a vector $\mathcal{V}$ with the length of $6+k$ was defined to characterize the 3D coordinates of the ligand:

$$\mathcal{V} = (x, y, z, \alpha, \beta, \gamma, \theta_1, \theta_2, \ldots, \theta_k) \qquad \text{Eq.10}$$

where the first 6 units refers to 3D coordinates $(x, y, z)$ of the first atom and the rotation angle $(\alpha, \beta, \gamma)$ of the molecule in space; $k$ and $\theta_k$ refers to the number of rotatable bonds and the angle of the $kth$ rotatable bond in the ligand molecule, respectively. For example, the ligand (PDB: 2WEG) shown in Fig. 1b contains two rotatable bonds that connect three substructures (identified by different colors). When the coordinates of the first atom in the first substructure and the rotation angle in space are provided, all positions of atoms in the first substructure can be determined according to the bonding relationship; then the positions of atoms in other substructures can be restored by twisting the rotatable bonds.

The docking pose was entered as a PDBQT file, which records the rotatable bonds of the ligand and divides the ligand molecule into multiple branches according to the rotatable bonds. The workflow of ligand pose optimization is shown in Fig. 1. First, the ligand molecule was encoded into a vector $\mathcal{V}$, which is the most original representation of the pose in the optimization process. Then, the 3D coordinates of the ligand ($R_{lig}$) were restored from the vector:

$$R_{lig} = \mathcal{M}(\mathcal{V}) \qquad \text{Eq.11}$$

which was used together with the 3D conformation of the protein ($R_{rec}$) for extracting the feature of the protein-ligand interaction and calculating the Vina score. The feature was processed and then fed into the MLP model to predict RMSD, which was summed with the Vina score as the combined score $\mathcal{L}(\mathcal{V})$:

$$\mathcal{L}(\mathcal{V}) = \mathcal{F}_{DeepRMSD}(R_{lig}, R_{rec}) + \mathcal{F}_{Vina}(R_{lig}, R_{rec}) \qquad \text{Eq.12}$$

where $\mathcal{F}_{DeepRMSD}$ and $\mathcal{F}_{Vina}$ represent the DeepRMSD and the Vina score, respectively. At this point, a computational loop with the ligand vector as input and the combined score as output was completed. To achieve iterative optimization of the pose, the gradient of the ligand vector in favor of the scoring function was calculated and applied to the current vector to generate a new ligand vector:

$$\nabla \mathcal{L} = (\frac{\partial \mathcal{L}}{\partial x}, \frac{\partial \mathcal{L}}{\partial y}, \dots, \frac{\partial \mathcal{L}}{\partial \theta_k}) \qquad \text{Eq.13}$$

$$\mathcal{V}' = \mathcal{V} - \eta \cdot \nabla \mathcal{L} \qquad \text{Eq.14}$$

where $\mathcal{V}'$ is updated ligand vector $\mathcal{V}$ and $\eta$ is an adjustable parameter. For each pose optimization task, 70 basic iterations were performed first, and then the early stopping strategy was introduced, which was similar to our MLP training. If the predicted combined score did not drop significantly over the next 30 iterations, the optimization process stopped and the conformation with the lowest predicted combined score was output. Fig. 1c shows a case of ligand docking pose optimization, and it is clear that the optimized pose closely approximates the native conformation.

In practical optimization scenarios, it is challenging to make all optimized poses higher quality than the initial poses. Therefore, we set up an optimization protocol: the optimized conformation is output only when the predicted value (predicted RMSD by DeepRMSD + Vina score) drops, otherwise the initial conformation is output.

## 3. Results and Discussion

### 3.1 DeepRMSD+Vina improves the docking power

In molecular docking, the SF is expected to distinguish near-native docking poses, which are defined as those with an RMSD less than 2 Å relative to the native pose. Usually, all SFs' output values are closely linked to binding affinity. The RMSD used here, however, can be considered as proxy for binding affinity. The hypothesis is that the pose with a smaller RMSD value has a greater binding affinity for the receptor.

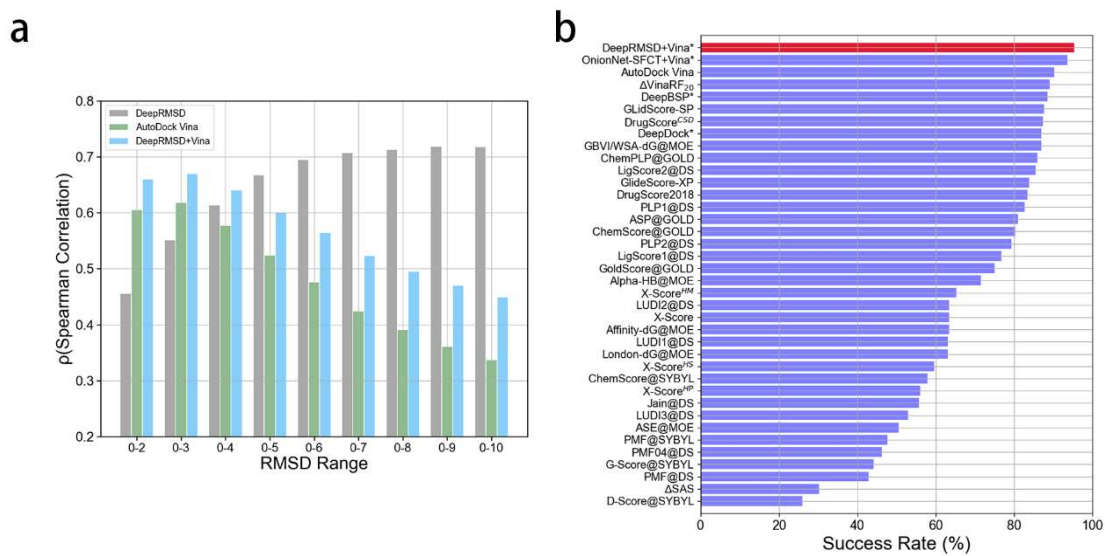

**Figure 2.** The performance of DeepRMSD (+Vina) on docking poses provided by CASF-2016. **a** Spearman correlation achieved by DeepRMSD, AutoDock Vina, and DeepRMSD+Vina on docking poses with different RMSD intervals. The output values of AutoDock Vina on CASF-2016 docking poses are derived from the article by Su et al[34]. **b** Comparison of DeepRMSD+Vina and other SFs on the docking power test with the Top 1 success rate as criterion. Two recent DL-based SFs DeepBSP and DeepDock, as well as our previously proposed OnionNet-SFCT+Vina (marked with the asterisk), are also included.

We first tested the performance of DeepRMSD and DeepRMSD+Vina on CASF-2016 docking poses. Fig. 2a shows the Spearman correlation coefficients between the output values achieved by DeepRMSD, AutoDock Vina, and DeepRMSD+Vina on CASF-2016 docking poses and their measured RMSD values. As the range of RMSD increased, the spearman correlation coefficient achieved by DeepRMSD showed an upward trend. Conversely, AutoDock Vina performed significantly better in the low RMSD interval than in the high RMSD interval. Since the DeepRMSD and Vina scores showed completely different trends depending on the RMSD interval, we tried combining these two algorithms with different weights to obtain a new SF (which we called DeepRMSD+Vina). Interestingly, the trends of DeepRMSD+Vina in different RMSD intervals were consistent with those of the Vina scores, but the Spearman correlation was significantly improved. Especially in the low RMSD range, DeepRMSD+Vina showed clear advantages compared to both DeepRMSD and AutoDock Vina, although DeepRMSD performed better in high RMSD intervals.

The main purpose of molecular docking is to identify one or several near-native docking poses, which usually appear in low RMSD intervals. Therefore, the accuracy of the SF in the low RMSD interval is critical for molecular docking performance. In view of this, we tested DeepRMSD+Vina against the docking power of CASF-2016 and compared with other SFs analyzed in CASF-2016 benchmark and two DL-based

SFs DeepBSP and DeepDock. The results are shown in Fig. 2b. Compared with other SFs, DeepRMSD+Vina showed great advantages, demonstrating a success rate of 95.4 % for Top 1, while the second-best SF was our previously proposed OnionNet-SFCT+Vina with 90.2 %. When the native binding poses are not included in the test set, DeepRMSD+Vina achieved the Top 1 success rate of 91.2 %, which was slightly better than DeepBSP (Supplementary Fig. 2).

**3.2 Constructing a ligand optimization framework based on DeepRMSD+Vina**

DeepRMSD+Vina is a SF entirely based on protein-ligand structure, which is differentiable for the coordinates of the ligand. For identical targets, the DeepRMSD+Vina score can be regarded as a physical quantity representing the strength of protein-ligand binding interaction. In view of the outstanding performance of DeepRMSD+Vina on the docking power test, we tried to build a ligand pose optimization framework using a gradient descent algorithm.

In this work, we employed the docking poses provided by CASF-2016 to test the performance of our optimization framework. We first defined an evaluation criterion: on the premise that as the new SF decreases the measured RMSD value also decreases, any given case was taken as a successful optimization case. The success rate of optimization ($sr$) was calculated by Eq. 15, where $N_{optimized\ cases}$ is the number of cases where the new SF decreases, and $N_{success\ cases}$ is the number of cases where the new SF decreases simultaneously with the measured RMSD.

$$sr(optimization) = \frac{N_{success\ cases}}{N_{optimized\ cases}} \qquad \text{Eq.15}$$

CASF-2016 docking poses were generated from 285 native protein-ligand complexes, each of which contained about 100 docking poses. Firstly, we calculated the optimization success rate of each target for 285 targets; the distribution of the number of targets occupying different success rate intervals is shown in Fig. 3a. Surprisingly, we observed that the optimization success rate was higher than 60 % for most target docking poses. Fig. 3b shows the optimization success rate for different RMSD intervals. As expected, the lower the initial RMSD of docking poses, the higher the success rate of pose optimization. Our optimization framework was based on DeepRMSD+Vina's calculation of the intermolecular interaction potential, thus the success rate of optimization depended on the accuracy of DeepRMSD+Vina. As shown in Fig. 2a, DeepRMSD+Vina performed significantly better on docking poses with low RMSD than that with high RMSD, which is consistent with the trend of the optimization success rate here.

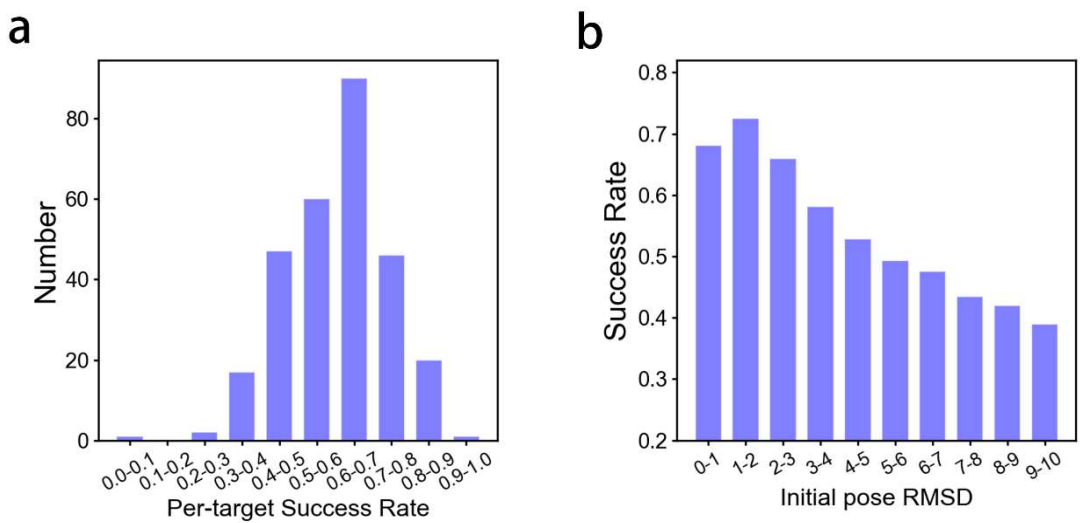

**Figure 3.** Results of optimization framework on CASF-2016 docking poses. **a** The distribution of the number of protein-ligand combinations in different success rate intervals. **b** The optimization success rate of DeepRMSD+Vina on each RMSD interval.

After optimization, the relationship between the change in RMSD and the number of active rotatable bonds and heavy atoms is shown in Supplementary Fig. 5 and Supplementary Fig. 6, respectively. For those poses with large RMSD values, the optimization algorithm had limited effects. One of the reasons is that for poses with high RMSD, they may need to translate, rotate, and twist considerably in space to return to the native conformation. However, these movements can be constrained by the size and shape of the target pocket due to spatial collision factors. In the context of the energy landscape, poses with high RMSDs are located far from the global minimum; the local structure optimization algorithm, such as the gradient descent algorithm used here, may not be able to traverse the energy barrier to locate the global minimum. More advanced global minimization algorithms should be applied in future work.

### 3.3 Evaluation of DeepRMSD+Vina and the optimization framework on the redocking and cross-docking tasks

To demonstrate the practicality of DeepRMSD+Vina and the optimization framework, we conducted additional tests on the redocking and cross-docking tasks.

In this test, the docking poses were fed into the optimization framework to obtain updated structures, and then DeepRMSD+Vina was used to score and rank these structures. The optimization success rates of DeepRMSD+Vina on the redocking and cross-docking sets are shown in Fig. 4a and 4c, respectively. The optimization success rate in the low RMSD intervals is significantly improved, which is generally consistent with the data presented in Fig. 3b. However, it cannot be ignored that the optimization success rate for docking poses with RMSD less than 1 Å in the redocking task is slightly decreased. One possibility is that the intermolecular potential energy functions (DeepRMSD+Vina) are not accurate enough to distinguish small deviations from the native pose. The docking performance as measured as Top N success rates on these two sets is presented in Fig. 4b and 4d. As controls, the performance of DeepRMSD and

AutoDock Vina on unoptimized structures were also reported. DeepRMSD and DeepRMSD+Vina both significantly outperform AutoDock Vina in both redocking and cross-docking tasks. As expected, DeepRMSD+Vina is robust in practical application scenarios. This indicates that our proposed framework for conformation optimization and selection has practical application value.

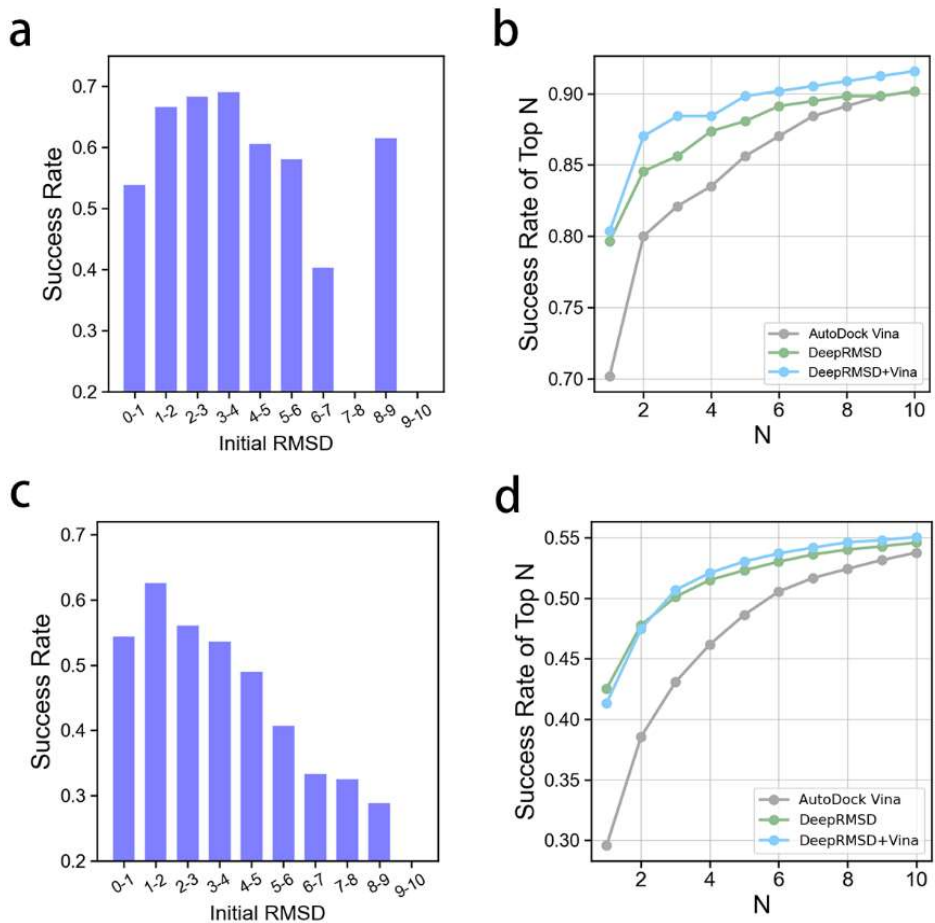

**Figure 4.** The performance of DeepRMSD+Vina on redocking and cross-docking tasks. **a** and **c** show the optimization success rate of DeepRMSD+Vina in different RMSD intervals on the redocking and cross-docking tasks, respectively. **b** and **d** show the performance of DeepRMSD+Vina on the optimized structures on the redocking and cross-docking tasks, respectively. As controls, the success rates of DeepRMSD and AutoDock Vina on unoptimized structures are also plotted.

**3.4 Analysis of protein-ligand interaction features**

The synergy exhibited by DeepRMSD and AutoDock Vina in low RMSD poses (Fig. 2a) indicates that the protein-ligand interaction information they describe are complementary. DeepRMSD describes the protein-ligand interaction through a vector of length 1470 containing 735 protein-ligand atom pairs. In order to evaluate the importance of features, we constructed a regression model based on a random forest algorithm in the scikit-learn package[44]. This random forest model is trained to fit the relationship between features and RMSD and then output the importance of each feature. As shown in Fig. 5, the top 50 most important features are mostly van der Waals interaction terms ($i = 6$). This suggests that the short-range interactions play a major role in protein-ligand binding. Furthermore, "VAL-C_C" and "LEU-C_C" were the top two most important pairs. "MET-S_C" is also an important combination, which is consistent with our previous study[29]. "VAL", "LEU" and "MET" all have hydrophobic side chains, which may indicate that the hydrophobic environment at the binding pocket is important for ligand binding. "ASP-O_N" is also recognized to be a combination pair that is particularly important to protein-ligand interaction, likely owing to the formation of hydrogen bonds.

Among the top 10 most important features, 60 % of the features indicated polar contacts. These polar contacts promote the specificity of the binding mode, and as such may contribute substantially to pose quality prediction.

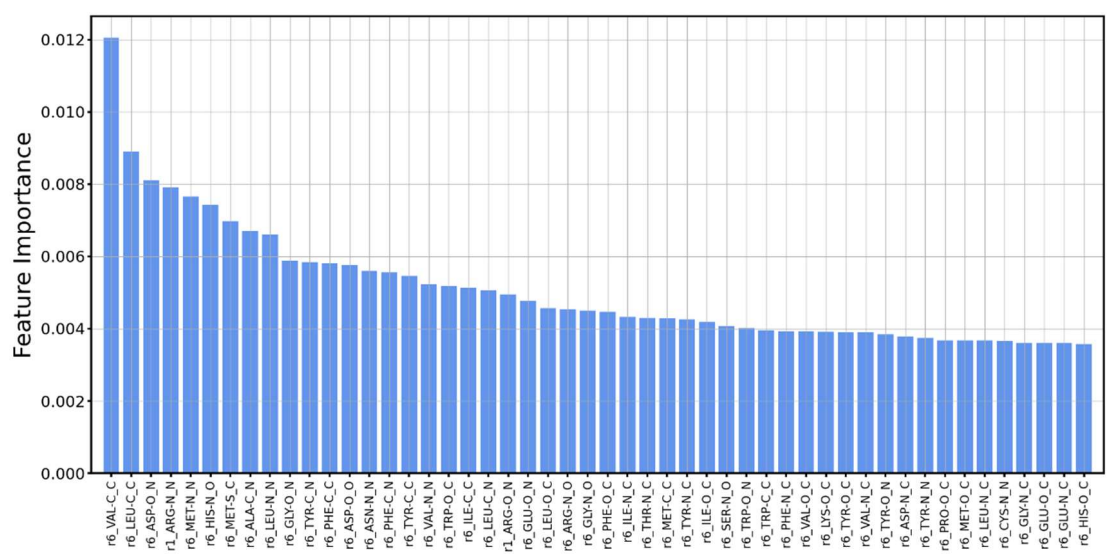

**Figure 5.** The importance of features derived from the random forest model. Only the top 50 most important features are shown.

Successfully optimized docking poses should possess stronger intermolecular interactions and should also be as close as possible to the native conformation. These high-quality poses with high binding affinity may be affected by specific intermolecular interactions during the binding process, such as hydrogen bonds, $\pi - \pi$ stacking, etc. Fig. 6 shows the hydrogen bond formation for a docking pose before and after optimization. The optimized pose clearly has more hydrogen bonds that contribute to protein-ligand binding. The animation of this optimization process can be seen in the Supplemental Material. DeepRMSD+Vina was built based on physical knowledge; thus it has the ability to discern critical physical interactions when optimizing ligand poses.

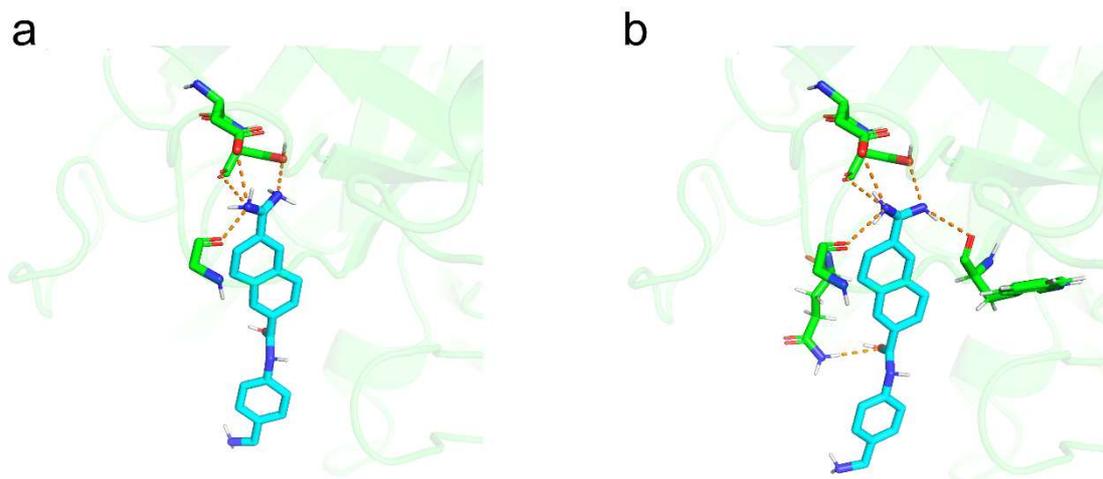

**Figure 6.** The hydrogen bond formation for a docking pose (PDB id: 1OWH) before (**a**) and after (**b**) optimization. The cyan stick represents the ligand conformation, and the green sticks represent the residues that form hydrogen bonds with the ligand. The yellow dotted lines represent hydrogen bonds.

### 3.5 Discussion for optimization framework

AutoDock Vina uses a quasi-Newton method called Broyden-Fletcher-Goldfarb-Shanno (BFGS) for local optimization, which takes into account not only the value of the scoring function but also the gradients of the scoring function with respect to its arguments[10], similarly to our optimization algorithm. Therefore, the initial poses of our optimization framework in redocking and cross-docking tasks are inherently optimized by AutoDock Vina. As shown in Fig. 4a and 4c, our optimization framework further improves the overall docking pose quality, especially in the low RMSD intervals. As a DL-based optimization algorithm with many more parameters than AutoDock Vina, the required computational time will necessarily be longer than that required for other optimization algorithms. The relationship between the optimization time and the number of heavy atoms in shown in Supplementary Fig. 6. It is worth noting that this issue will be resolved with the introduction of graphical processing units (GPUs).

Protein-ligand binding affinity is heavily dependent on binding pose. Currently, the free energy perturbation (FEP) based on molecular dynamics (MD) simulation can yield accurate predictions of protein-ligand binding affinity[7]. However, the prerequisite for accurate calculation is that a reliable protein-ligand complex structure must be provided. The poses of a ligand in the target pocket in virtual screening are usually generated by molecular docking programs, and their quality still needs to be improved. Currently, the optimization of ligand conformations by deep learning algorithms remains to be explored, and the SFs differentiable to molecular coordinates will have greater application value. We constructed a ligand conformation optimization framework based on DeepRMSD+Vina, which can improve the quality of binding poses by further conformation optimization, thereby improving the accuracy of binding affinity prediction.

In addition, it can be seen in Fig. 4b and 4d that DeepRMSD+Vina can achieve higher docking success rates with our optimization framework. Therefore, our current work establishes the feasibility of implementing a highly accurate molecular software that significantly outperforms AutoDock Vina.

**3.6 In comparison with other ML/DL models targeting at RMSD prediction**

The majority of currently available ML/DL models dealing with protein-ligand interaction are tailored to binding affinity prediction. Very limited validated experimental binding affinity data exists, however, and this information is largely associated with native poses. The training sets for these models are therefore quite limited. On the other hand, in real virtual screening settings, a variety docking pose decoys and molecule decoys are relied upon. It is therefore unsurprising that those DL models (including Pafnucy[46] and our previous models, and OnionNet[29]) which perform quite well in true binder molecule binding affinity prediction can hardly distinguish molecule decoys from the true binders[35].

RMSD calculations only require information about native poses; experimental binding affinity data is not critical to these calculations. More importantly, they provide objective labels for every docking pose. This allows for a training set that is one or two orders of magnitude larger than the dataset limited only to native poses. Recently, the DeepBSP model applied 3D CNN to predict RMSD[47]. Using protein-ligand complexes' 3D structures with voxel grids as input features, the model achieved an average Top 1 success rate of 88.5 % benchmarked in CASF-2016 docking power, while DeepDock[40] achieved the Top 1 success rate of 87 %. Our DeepRMSD+Vina model can achieve a Top 1 success rate of 95.4 % benchmarked in the same dataset. Utilizing 3D complex structures as features seems to provide a "full" picture of protein-ligand interaction. We then rely on a simple MLP model to learn the "underlying physics" of these interactions to infer the nature of the binding process based on comprehensive structural details. The physical-based features used in our model, the $\frac{1}{r}$, and $\frac{1}{r^6}$ terms, represent the key variables controlling the binding/recognition processes. Feeding them into the MLP simplifies the learning process and leads to robust outputs.

## Conclusion

In this work, DeepRMSD, a MLP model for predicting the RMSD of docking poses of ligands relative to native conformation, is proposed. Inspired by physical knowledge, DeepRMSD extracts protein-ligand interaction features based on van der Waals and electrostatic interactions. When combined with Vina score, the model will significantly improve the docking power in CASF-2016 benchmark. In view of this, we designed a pose optimization framework based on DeepRMSD+Vina. The optimization algorithm can achieve high optimization success rates on CASF-2016 docking poses with low RMSD. In redocking and cross-docking tasks, applying the DeepRMSD+Vina algorithm will effectively improve the success rate of near-native pose screening. This demonstrates the high practical application value of an optimization framework based on DeepRMSD+Vina.

**Acknowledgement**

This work is supported by the Natural Science Foundation of Shandong Province (ZR2020JQ04), and National Natural Science Foundation of China (11874238).

This work is partly supported by the National Key Research and Development Program of China under Grant No. 2018YFB0204403, Strategic Priority CAS Project XDB38050100, the Key Research and Development Project of Guangdong Province under grant no. 2021B0101310002, National Science Foundation of China under grant no. U1813203, the Shenzhen Basic Research Fund under grant no RCYX2020071411473419, JCYJ20200109114818703 and JSGG20201102163800001, CAS Key Lab under grant no. 2011DP173015. We would also like to thank the funding support by the Youth Innovation Promotion Association(Y2021101), CAS to Yanjie Wei.

This work also is supported by Singapore MOE Tier 1 grant RG27/21.